\documentstyle{article}
\newtheorem{theo}{Theorem}

\def\ds{\displaystyle}
\def\beq{\begin{equation}}
\def\eeq{\end{equation}}
\def\bea{\begin{eqnarray}}
\def\eea{\end{eqnarray}}
\def\nn{\nonumber}
\def\deg{\mathop{\rm deg}\nolimits}
\def\tr{\mathop{\rm tr}\nolimits}
\def\str{\mathop{\rm str}\nolimits}
\def\Res{\mathop{\rm Res}\nolimits}
\def\qdots{\mathinner{\mkern1mu\raise1pt\vbox{\kern7pt\hbox{.}}\mkern2mu
 \raise4pt\hbox{.}\mkern2mu\raise7pt\hbox{.}\mkern1mu}}
\def\al{\alpha}
\def\be{\beta}
\def\de{\delta}
\def\ep{\epsilon}  \def\vep{\varepsilon}
\def\th{\theta}

\def\rh{\rho}
\def\De{\Delta}
\def\La{\Lambda}
\def\Zah{{\bf Z}}
\def\C{{\bf C}}
\begin{document}
\addtolength{\baselineskip}{1mm}
\begin{center}
{\LARGE
Finite-dimensional representations of the quantum superalgebra
$U_q[gl(n/m)]$ and related $q$-identities
}\\[2cm]
{\Large T.D.~Palev$^{*,a,b,c)}$, N.I.~Stoilova$^{b,c)}$ and
J.~Van der Jeugt$^{\dagger,b)}$} \\[1cm]
\end{center}
\noindent
$^{a)}$ International Centre for Theoretical
Physics, 34100 Trieste, Italy.\\[5mm]
$^{b)}$ Department of Applied Mathematics and Computer Science,
University of Ghent, Krijgslaan 281--S9, B-9000 Gent, Belgium.\\[5mm]
$^{c)}$ Institute for Nuclear Research and
Nuclear Energy, 1784 Sofia, Bulgaria (permanent address).
\vspace{3cm}
\begin{abstract}
{\normalsize
Explicit expressions for the generators of the quantum superalgebra
$U_q[gl(n/m)]$ acting on a class of irreducible representations are given.
The class under consideration consists of all essentially typical
representations~: for these a Gel'fand-Zetlin basis is known.
The verification of the quantum superalgebra relations to be satisfied
is shown to reduce to a set of $q$-number identities.
}
\end{abstract}
\vspace{1cm}
\noindent
MSC numbers: 16W30, 17B40, 81R50.
\vfill
\noindent-----------------------------------\\
$^{*)}$ {\footnotesize E-mail~: palev@bgearn.bitnet.}\\
$^{\dagger)}$ {\footnotesize Research Associate of the N.F.W.O.
(National Fund for Scientific Research of Belgium). E-mail~:
joris@fwet.rug.ac.be.}
\newpage
\noindent
{\LARGE
Finite-dimensional representations of $U_q[gl(n/m)]$
}\\[2cm]

\section{Introduction}

This paper is devoted to the study of a class of
finite-dimensional irreducible representations of the quantum
superalgebra $U_q[gl(n/m)]$. The main goal is to present
explicit actions of the $U_q[gl(n/m)]$ generating elements
acting on a Gel'fand-Zetlin-like basis, and to discuss some
of the $q$-number identities related to these representations.

Quantum groups~\cite{drinfeld}, finding their origin in
the quantum inverse problem method~\cite{faddeev} and in investigations
related to the Yang-Baxter equation~\cite{jimbo90},
have now become an important and widely used concept in various
branches of physics and mathematics. A quantum (super)algebra $U_q[G]$
associated with a (simple) Lie (super)algebra $G$ is a deformation of
the universal enveloping algebra of $G$ endowed with a Hopf algebra
structure. The first example was given in~\cite{kulish-r80,sklyanin},
and soon followed the generalization to any Kac-Moody Lie algebra
with symmtrizable Cartan matrix~\cite{drinfeld85,jimbo85}.
For the deformation of the enveloping algebra of a Lie
superalgebra we mention the case of
$osp(1/2)$~\cite{kulish,kulish-r89},
later to be extended to Lie superalgebras with a symmetrizable
Cartan matrix~\cite{tolstoy} including the basic~\cite{kac77} Lie
superalgebras~\cite{bracken-g-z,chaichian-k}.

Representations of quantum algebras have been studied extensively,
particularly for generic $q$-values (i.e.~$q$ not a root of unity).
In this case, finite-dimensional irreducible representations
of $sl(n)$ can be deformed into irreducible representations of
$U_q[sl(n)]$~\cite{jimbo86}, and it was shown that one obtains
all finite-dimensional irreducible modules of $U_q[sl(n)]$
in this way~\cite{rosso}. In~\cite{jimbo}, explicit expressions
for the generators of $U_q[sl(n)]$ acting on the ``undeformed''
Gel'fand-Zetlin basis were given. It is in the spirit of this
work that our present paper should be seen. Here, we study
a class of irreducible representations of the quantum superalgebra
$U_q[gl(n/m)]$. The class consists of so-called essentially
typical representations; one can interpret these as irreducible
representations for which a Gel'fand-Zetlin basis can be given.
Just as in the case of $sl(n)$, the basis will remain undeformed,
and the deformation will arise in the action of the quantum
superalgebra generators on the basis vectors.

In the literature, some results have already appeared for representations of
quantum superalgebras of type $U_q[gl(n/m)]$~: representations
of $U_q[gl(n/1)]$ (both typical and atypical)
were examined in~\cite{palev-t} following
the study of~\cite{palev87,palev89-1}; a
generic example, $U_q[gl(3/2)]$, was treated in~\cite{palev-s};
and the induced module construction of Kac~\cite{kac79} was
generalized to $U_q[gl(n/m)]$~\cite{zhang}.
On the other hand, oscillator representations have been
constructed~\cite{chaichian-k,floreanini-s-v90,floreanini-s-v91}
not only for $U_q[gl(n/m)]$ but also for other quantum superalgebras.

The structure of the present paper is as follows. In Section~2
we recall the definition of the Lie superalgebra $gl(n/m)$ and
fix the notation. We also remind the reader of some representation
theory of $gl(n/m)$ which will be needed in the case of $U_q[gl(n/m)]$,
in particular of the concept of typical,
atypical, and essentially typical representations. For the
last class of representations, the Gel'fand-Zetlin basis
introduced in~\cite{palev89-2} is written in explicit form.
In the next section, we briefly recall the definition of
the quantum superalgebra $U_q[gl(n/m)]$. Section~4 contains our
main results. We present the actions of the $U_q[gl(n/m)]$
generators on the Gel'fand-Zetlin basis introduced, and we
give some indications of how the relations were proved in these
representations. Some of the relations actually reduce to
interesting $q$-number identities, which can be proved using
the Residue theorem of complex analysis. We conclude the paper
with some comments and further outlook.

\section{The Lie superalgebra $gl(n/m)$ and Gel'fand-Zetlin patterns}

The Lie superalgebra $G=gl(n/m)$ can be
defined~\cite{kac77,scheunert79} through its natural
matrix realization
\beq
gl(n/m)=\{ x=\left(\begin{array}{cc} A&B\\C&D\end{array}\right)
| A\in M_{n\times n}, B\in M_{n\times m}, C\in M_{m\times n},
  D\in M_{m\times m} \},
\label{defgl}
\eeq
where $M_{p\times q}$ is the space of all $p\times q$ complex matrices.
The even subalgebra $gl(n/m)_{\bar 0}$ has $B=0$ and $C=0$; the odd
subspace $gl(n/m)_{\bar 1}$ has $A=0$ and $D=0$. The bracket is determined
by
\beq
[a,b]=ab-(-1)^{\al\be}ba,\qquad\forall a\in G_\al\hbox{ and }
\forall b\in G_\be,
\eeq
where $\al,\be\in\{ \bar 0, \bar 1 \} \equiv \Zah_2$. If $a\in G_\al$
then $\al=\deg(a)$ is called the degree of $a$, and an element of
$G=G_{\bar 0}\oplus G_{\bar 1}$ is called homogeneous if it belongs
to either $G_{\bar 0}$ or else $G_{\bar 1}$.
We denote by $gl(n/m)_{+1}$ the space of matrices
$\left(\begin{array}{cc}0&B\\0&0\end{array}\right)$ and by
$gl(n/m)_{-1}$ the space of matrices
$\left(\begin{array}{cc}0&0\\C&0\end{array}\right)$. Then
$G=gl(n/m)$ has a $\Zah$-grading which is consistent with the
$\Zah_2$-grading~\cite{scheunert79}, namely $G=G_{-1}\oplus G_0\oplus
G_{+1}$ with $G_{\bar 0}=G_0$ and $G_{\bar 1}=G_{-1}\oplus G_{+1}$.
Note that $gl(n/m)_0 = gl(n)\oplus gl(m)$. For elements $x$ of
$gl(n/m)$ given by (\ref{defgl}), one defines the supertrace
as $\str(x)=\tr(A)-tr(D)$. The Lie superalgebra $gl(n/m)$ is
not simple, and (for $n\ne m$) one can define the simple superalgebra $sl(n/m)$
as the subalgebra consisting of elements with supertrace $0$. However,
the representation theory of $gl(n/m)$ or $sl(n/m)$ is essentially
the same (the situation is similar as for the classical Lie
algebras $gl(n)$ and $sl(n)$), and hence we prefer to work with
$gl(n/m)$ and in the following section with its Hopf superalgebra
deformation $U_q[gl(n/m)]$.

A basis for $G=gl(n/m)$ consists of matrices $E_{ij}$ ($i,j=1,2,\ldots,
r\equiv m+n$) with entry $1$ at position $(i,j)$ and $0$ elsewhere.
A Cartan subalgebra $H$ of $G$ is spanned by the elements $h_j=E_{jj}$
($j=1,2,\ldots,r$), and a set of generators of $gl(n/m)$ is given
by the $h_j$ ($j=1,\ldots,r$) and the elements $e_i=E_{i,i+1}$ and
$f_i=E_{i+1,i}$ ($i=1,\ldots,r-1$). The space dual to $H$ is $H^*$ and
is described by the forms
$\ep_i$ ($i=1,\ldots,r$) where
$\ep_j:x\rightarrow A_{jj}$ for $1\leq j\leq n$ and $\ep_{n+j}:x\rightarrow
D_{jj}$ for $1\leq j\leq m$, and where $x$ is
given as in (\ref{defgl}). On $H^*$ there is a bilinear form
defined deduced from the supertrace on $G$, and explicitly given
by~\cite{vdjeugt-h-k-t2}:
\beq
\begin{array}{ll}
\langle \ep_i|\ep_j \rangle =\de_{ij},&\hbox{ for } 1\leq i,j \leq n;\\
\langle \ep_i|\ep_p \rangle =0,&\hbox{ for } 1\leq i \leq n\hbox{ and }
 n+1\leq p \leq r;\\
\langle \ep_p|\ep_q \rangle =-\de_{pq},&\hbox{ for } n+1\leq p,q\leq r.
\end{array}
\eeq
where $\de_{ij}$ is the Kronecker-$\de$. The components of an
element $\La\in H^*$ will be written as $[m]=[m_{1r},m_{2r},\ldots,m_{rr}]$
where $\La=\sum_{i=1}^r m_{ir}\ep_i$ and $m_{ir}$ are complex numbers.
The elements of $H^*$ are called the weights. The roots of $gl(n/m)$
are the non-zero weights of the adjoint representation, and take the
form $\ep_i-\ep_j$ ($i\ne j$) in this basis; the positive roots are
those with $1\leq i<j\leq r$, and of importance are the $nm$ odd positive roots
\beq
\be_{ip} = \ep_i-\ep_p, \quad\hbox{ with }1\leq i\leq n\hbox{ and }
 n+1\leq p \leq r.
\label{odd}
\eeq
For an element $\La\in H^*$ with components $[m]$, the Kac-Dynkin labels
$(a_1,\ldots,a_{n-1};a_n;a_{n+1},\ldots,a_{r-1})$ are given by
$a_i=m_{ir}-m_{i+1,r}$ for $i\ne n$ and $a_n=m_{nr}+m_{n+1,r}$.
Hence, $\La$ with components $[m]$ will be called an integral dominant
weight if $m_{ir}-m_{i+1,r}\in\Zah_+=\{0,1,2,\ldots\}$ for all $i\ne n$
($1\leq i\leq r-1$). For every integral dominant weight $\La\equiv[m]$
we denote by $V^0(\La)$ the simple $G_0$ module with highest weight
$\La$; this is simply the finite-dimensional
$gl(n)\oplus gl(m)$ module with $gl(n)$
labels $\{m_{1,r},\ldots m_{nr}\}$ and with $gl(m)$ labels
$\{m_{n+1,r},\ldots,m_{rr}\}$. The module $V^0(\La)$ can be extended
to a $G_0\oplus G_{+1}$ module by the requirement that $G_{+1}V^0(\La)=0$.
The induced $G$ module $\overline V(\La)$, first introduced by
Kac~\cite{kac79} and usually referred to as the Kac-module, is defined
by
\beq
\overline V(\La) = \hbox{Ind}_{G_0\oplus G_{+1}}^{G} V^0(\La)
 \cong U(G_{-1})\otimes V^0(\La),
\eeq
where $U(G_{-1})$ is the universal enveloping algebra of $G_{-1}$.
It follows that $\dim \overline V(\La) = 2^{nm} \dim V^0(\La)$.
By definition, $\overline V(\La)$ is a highest weight module;
unfortunately, $\overline V(\La)$ is not always a simple $G$ module.
It contains a unique maximal (proper) submodule $M(\La)$, and the quotient
module
\beq
V(\La)=\overline V(\La)/M(\La) \label{vl}
\eeq
is a finite-dimensional
simple module with highest weight $\La$. In fact, Kac~\cite{kac79}
proved the following~:
\begin{theo} Every finite-dimensional simple $G$ module is isomorphic
to a module of type~(\ref{vl}), where $\La\equiv[m]$ is integral dominant.
Moreover, every finite-dimensional simple $G$ module is uniquely
characterized by its integral dominant highest weight $\La$.
\end{theo}
An integral dominant weight $\La=[m]$ (resp.~$\overline V(\La)$,
resp.~$V(\La)$) is called a typical weight (resp.~a typical Kac
module, resp.~a typical simple module) if and only if $\langle
\La+\rh|\be_{ip}\rangle\ne 0$ for all odd positive roots
$\be_{ip}$ of (\ref{odd}),
where $2\rh$ is the sum of all positive roots of $G$. Otherwise
$\La$, $\overline V(\La)$ and $V(\La)$ are called atypical.
The importance of these definitions follows from another theorem of
Kac~\cite{kac79}~:
\begin{theo}
The Kac-module $\overline V(\La)$ is a simple $G$ module if and only
if $\La$ is typical.
\end{theo}
For an integral dominant highest weight $\La=[m]$ it is convenient
to introduce the following labels~\cite{palev89-2}~:
\beq
l_{ir}=m_{ir}-i+n+1,\quad(1\leq i \leq n);\qquad
l_{pr}=-m_{pr}+p-n,\quad(n+1\leq p\leq r).
\label{lir}
\eeq
In terms of these, one can deduce that $\langle\La+\rh|\be_{ip}\rangle=
l_{ir}-l_{pr}$, and hence the conditions for typicality take
a simple form.

For typical modules or representations one can say that they are
well understood, and a character formula was given by Kac~\cite{kac79}.
A character formula for all atypical modules has not been proven so
far, but there are several breakthroughs in this area~: for singly atypical
modules (for which the highest weight $\La$ is atypical with respect
to one single odd root $\be_{ip}$) a formula has been
constructed~\cite{vdjeugt-h-k-t2}; for all atypical modules a
formula has been conjectured~\cite{vdjeugt-h-k-t1}; for atypical
Kac-modules the composition series has been conjectured~\cite{hughes-k-v}
and partially shown to be correct~\cite{su-h-k-v}.
On the other hand, the modules for which an explicit action of generators on
basis vectors can be given, similar to the action of generators
of $gl(n)$ on basis vectors with Gel'fand-Zetlin labels, is only
a subclass of the typical modules, namely the so-called essentially
typical modules~\cite{palev89-2}, the definition of which shall
be recalled here.

For simple $gl(n)$ modules the Gel'fand-Zetlin basis
vectors~\cite{gelfand-z} and
their labels -- with the conditions (``in-betweenness conditions'') --
are reflecting the decomposition of the module with respect to
the chain of subalgebras $gl(n)\supset gl(n-1)\supset \cdots \supset gl(1)$.
In trying to construct a similar basis for the finite-dimensional modules
of the Lie superalgebra $gl(n/m)$ it was natural to consider the
decomposition with respect to the chain of subalgebras $gl(n/m)
\supset gl(n/m-1) \supset \cdots \supset gl(n/1) \supset gl(n)
\supset gl(n-1) \supset \cdots \supset gl(1)$. However, in order
to be able to define appropriate actions of the $gl(n/m)$ generators
on basis vectors with respect to this decomposition, it was necessary
that at every step in this reduction the corresponding modules are
completely reducible with respect to the submodule under consideration.
A sufficient condition is that for every step in the above reduction
the modules are typical, i.e.~a typical $gl(n/m)$ module $V$ must
decompose into typical $gl(n/m-1)$ modules, each of which must
decompose into typical $gl(n/m-2)$ modules etc. Such modules are
called essentially typical~\cite{palev89-2}, and a Gel'fand-Zetlin-like
basis can be constructed with an action of the $gl(m/n)$ generators.
In terms of the above quantities $l_{ir}$, a module with highest
weight $\La\equiv[m]$ is essentially typical if and only if
\beq
\{l_{1r},l_{2r},\ldots,l_{nr}\} \cap
\{l_{n+1,r},l_{n+1,r}+1,l_{n+1,r}+2,\ldots,l_{rr}\}=\emptyset.
\label{esstyp}
\eeq
The explicit form of the action~\cite{palev89-1,palev89-2} will not
be repeated here,
but the reader interested can deduce it from relations
(\ref{ki}--\ref{fp}) of the present paper by taking the limit
$q\rightarrow 1$ (in fact, the limit of our present relations also
improve some minor misprints in the transformations of the GZ basis
as given in~\cite{palev89-1,palev89-2}).
It is necessary, however, to recall the labelling
of the basis vectors for these modules, since the labelling of basis
vectors of representations of the quantum algebra $U_q[gl(n/m)]$ is
exactly the same (note that also for the quantum algebra $U_q[gl(n)]$,
the finite-dimensional representations can be labelled by the
same Gel'fand-Zetlin patterns as in the non-deformed case of $gl(n)$,
when $q$ is not a root of unity~\cite{jimbo}).

Let $[m]$ be the labels of an integral dominant weight $\La$.
Associated with $[m]$ we define a pattern $|m)$ of $r(r+1)/2$ complex
numbers $m_{ij}$ ($1\leq i\leq j\leq r$) ordered as in the usual
Gel'fand-Zetlin basis for $gl(r)$~:
\beq
|m) = \left|
\begin{array}{lclllcll}
m_{1r} & \cdots & m_{n-1,r} & m_{nr} & m_{n+1,r} & \cdots & m_{r-1,r}
& m_{rr}\\
m_{1,r-1} & \cdots & m_{n-1,r-1} & m_{n,r-1} & m_{n+1,r-1} & \cdots
& m_{r-1,r-1} & \\
\vdots & \vdots &\vdots &\vdots & \vdots & \qdots & & \\
m_{1,n+1} & \cdots & m_{n-1,n+1} & m_{n,n+1} & m_{n+1,n+1} & & & \\
m_{1n} & \cdots & m_{n-1,n} & m_{nn} & & & & \\
m_{1,n-1} & \cdots & m_{n-1,n-1} & & & & & \\
\vdots & \qdots & & & & & & \\
m_{11} & & & & & & &
\end{array}
\right)
\label{m}
\eeq
Such a pattern should satisfy the following set of conditions~:
\beq
 \begin{array}{rl}
1.& \hbox{the labels }m_{ir}\hbox{ of }\La\hbox{ are fixed for all patterns,}\\
2.& m_{ip}-m_{i,p-1}\equiv\th_{i,p-1}\in\{0,1\},\quad(1\leq i\leq n;\;
    n+1\leq p\leq r),\\
3.& m_{ip}-m_{i+1,p}\in\Zah_+,\quad(1\leq i\leq n-1;\;
    n+1\leq p\leq r),\\
4.& m_{i,j+1}-m_{ij}\in\Zah_+\hbox{ and }m_{i,j}-m_{i+1,j+1}\in\Zah_+,\quad
    (1\leq i\leq j\leq n-1\hbox{ or } n+1\leq i\leq j\leq r-1).
 \end{array}
\label{cond}
\eeq
The last condition corresponds to the in-betweenness condition and
ensures that the triangular pattern to the right of the $m\times n$
rectangle $m_{ip}$ ($1\leq i\leq n$; $n+1\leq p\leq r$) in (\ref{m})
corresponds to a classical Gel'fand-Zetlin pattern for $gl(m)$,
and that the triangular pattern below this rectangle corresponds
to a Gel'fand-Zetlin pattern for $gl(n)$.

The following theorem was proved~\cite{palev89-2}~:
\begin{theo}
Let $\La\equiv[m]$ be an essentially typical highest weight.
Then the set of all patterns~(\ref{m}) satisfying~(\ref{cond})
constitute a basis for the (typical) Kac-module $\overline V(\La)=V(\La)$.
\end{theo}
The patterns~(\ref{m}) are referred to as Gel'fand-Zetlin (GZ) basis
vectors for $V(\La)$ and an explicit action of the $gl(n/m)$
generators $h_j$ ($1\leq j\leq r$), $e_i$ and $f_i$ ($1\leq i \leq r-1$)
has been given in Ref.~\cite{palev89-2}.

In the following section we shall recall the definition of the
quantum algebra $U_q[gl(n/m)]$. We shall then define an action of
the quantum algebra generators on the Gel'fand-Zetlin basis
vectors $|m)$ introduced here. In other words, just as for the
finite-dimensional $gl(n)$ modules, one can use the same basis
vectors and only the action is deformed.

\section{The quantum superalgebra $U_q[gl(n/m)]$}

The quantum superalgebra $U_q\equiv U_q[gl(n/m)]$ is the free associative
superalgebra over $\C$ with parameter $q\in\C$ and
generators $k_j$, $k_j^{-1}$ ($j=1,2,\ldots,r\equiv n+m$)
and $e_i$, $f_i$ ($i=1,2,\ldots,r-1$) subject to the following
relations (unless stated otherwise, the indices below run over
all possible values)~:
\begin{itemize}
\item The Cartan-Kac relations~:
\beq
k_ik_j=k_jk_i,\qquad k_ik_i^{-1}=k_i^{-1}k_i=1;  \label{kk}
\eeq
\beq
k_ie_jk_i^{-1}=q^{(\delta_{ij}-\delta_{i,j+1})/2}e_j,
\qquad
k_if_jk_i^{-1}=q^{-(\delta_{ij}-\delta_{i,j+1})/2}f_j;
 \label{kek}
\eeq
\bea
e_if_j-f_je_i&=&0 \ \hbox{ if } i\neq j; \label{eifj} \\
e_if_i-f_ie_i&=&(k_i^2k_{i+1}^{-2}-k_{i+1}^2k_i^{-2})/(q-q^{-1})
 \ \hbox{ if } i\ne n; \label{eifi}\\
e_nf_n+f_ne_n&=&(k_n^2k_{n+1}^2-k_n^{-2}k_{n+1}^{-2})/(q-q^{-1}) ; \label{enfn}
\eea
\item The Serre relations for the $e_i$ ($e$-Serre relations)~:
\beq
e_ie_j=e_je_i \hbox{ if } \vert i-j \vert \neq 1;\qquad
e_n^2=0; \label{ee}
\eeq
\beq
e_i^2e_{i+1}-(q+q^{-1})e_ie_{i+1}e_i+e_{i+1}e_i^2=0,
\hbox{ for } i\in \{1,\ldots,n-1\}\cup\{n+1,\ldots,n+m-2\}; \label{eee1}
\eeq
\beq
e_{i+1}^2e_i-(q+q^{-1})e_{i+1}e_ie_{i+1}+e_ie_{i+1}^2=0,
\hbox{ for } i\in\{1,\ldots,n-2\}\cup\{n,\ldots,n+m-2\}; \label{eee2}
\eeq
\beq
e_ne_{n-1}e_ne_{n+1}+e_{n-1}e_ne_{n+1}e_n+e_ne_{n+1}e_ne_{n-1}
 +e_{n+1}e_ne_{n-1}e_n-(q+q^{-1})e_ne_{n-1}e_{n+1}e_n=0;
 \label{eeee}
\eeq
\item The relations obtained from (\ref{ee}--\ref{eeee}) by replacing
every $e_i$ by $f_i$ ($f$-Serre relations).
\end{itemize}

Equation~(\ref{eeee}) is the so-called extra Serre
relation~\cite{floreanini-l-v,khoroshkin-t,scheunert}.
The $\Zah_2$-grading in $U_q$ is defined by the requirement that
the only odd generators are $e_n$ and $f_n$; the degree of a
homogeneous element $a$ of $U_q$ shall be denoted by $\deg(a)$.
It can be shown that
$U_q$ is a Hopf superalgebra with counit $\vep$, comultiplication $\De$
and antipode $S$, defined by~:
\beq
\vep(e_i)=\vep(f_i)=0,\qquad \vep(k_j)=1; \label{vep}
\eeq
\beq
 \begin{array}{llll}
 \Delta(k_j)&=&k_j \otimes k_j,& \\
 \Delta(e_i)&=&e_i \otimes k_ik_{i+1}^{-1} + k_i^{-1}k_{i+1} \otimes e_i,
  &\hbox{ if } i\ne n, \\
 \Delta(e_n)&=&e_n \otimes k_nk_{n+1} + k_n^{-1}k_{n+1}^{-1} \otimes e_n, & \\
 \Delta(f_i)&=&f_i \otimes k_ik_{i+1}^{-1} + k_i^{-1}k_{i+1} \otimes f_i,
  &\hbox{ if } i\neq n, \\
 \Delta(f_n)&=&f_n \otimes k_nk_{n+1} + k_n^{-1}k_{n+1}^{-1} \otimes f_n; &
 \end{array}
\label{De}
\eeq
\beq
 \begin{array}{llll}
 S(k_j)&=&k_j^{-1}, & \\
 S(e_i)&=&-qe_i,\quad S(f_i)=-q^{-1}f_i,& \hbox{ if }i\ne n,\\
 S(e_n)&=&-e_n,\quad S(f_n)=-f_n. &
 \end{array}
\label{S}
\eeq
Remember that $\De : U_q\rightarrow U_q\otimes U_q$ is a morphism of
{\em graded} algebras, and that the multiplication in $U_q\otimes U_q$
is given by
\beq
(a\otimes b)(c\otimes d)= (-1)^{\deg(b)\deg(c)} ac\otimes bd.
\eeq

\section{The $U_q[gl(n/m)]$ representations}

Let $\La\equiv[m]$ be an essentially typical highest weight, and
denote by $W(\La)$ the vector space spanned by the basis vectors
$|m)$ of the form~(\ref{m}) satisfying the conditions~(\ref{cond}).
On this vector space, we shall define an action of the generators
of $U_q=U_q[gl(n/m)]$, thus turning $W(\La)$ into a $U_q$ module.
For convenience, we introduce the following notations~: $l_{ij}=
m_{ij}-i+n+1$ for $1\leq i\leq n$, $l_{pj}=-m_{pj}+p-n$ for
$n+1\leq p\leq r$, and $|m)_{\pm ij}$ is the pattern obtained
from $|m)$ by replacing the entry $m_{ij}$ by $m_{ij}\pm 1$.

The following is the main result of this paper (as usual, $[x]$ stands
for $(q^x-q^{-x})/(q-q^{-1})$)~:
\begin{theo}
For generic values of $q$ every essentially typical
$gl(n/m)$ module $V(\La)$ with highest weight $\La$
can be deformed into an irreducible $U_q[gl(n/m)]$ module $W(\La)$ with
the same underlying vector space and with the action of the generators
given by~:
\end{theo}
\bea
k_i|m)&=&q^{\left(\sum_{j=1}^i m_{ji}-
\sum_{j=1}^{i-1} m_{j,i-1}\right)/2 }|m), \quad
(1\leq i\leq r),\label{ki}\\
e_k|m)&=&\sum_{j=1}^k \left(-
{\prod_{i=1}^{k+1} [l_{i,k+1}-l_{jk}]
\prod_{i=1}^{k-1} [l_{i,k-1}-l_{jk}-1]
\over \prod_{i\neq j=1}^k [l_{ik}-l_{jk}]
[l_{ik}-l_{jk}-1] } \right)^{1/2}|m)_{jk},\quad (1\leq k\leq n-1),\label{ek}\\
f_k|m)&=&\sum_{j=1}^k \left(-
{\prod_{i=1}^{k+1} [  l_{i,k+1}-l_{jk}+1  ]
\prod_{i=1}^{k-1} [  l_{i,k-1}-l_{jk}  ]
\over \prod_{i\neq j=1}^k [  l_{ik}-l_{jk}+1  ]
[  l_{ik}-l_{jk}  ] } \right)^{1/2}|m)_{-jk},\quad (1\leq k\leq n-1),
\label{fk}\\
e_n|m)&=&\sum_{i=1}^n \theta_{in}(-1)^{i-1}
(-1)^{\theta_{1n}+ \ldots +\theta_{i-1,n} }
\left(\prod_{k=1}^{n-1} [  l_{k,n-1}-l_{i,n+1} ]
\over \prod_{k\neq i=1}^n [  l_{k,n+1}-l_{i,n+1} ]
\right)^{1/2} |m)_{in},\label{en}\\
f_n|m)&=&\sum_{i=1}^n (1-\theta_{in})(-1)^{i-1}
(-1)^{\theta_{1n}+ \ldots +\theta_{i-1,n} }[l_{i,n+1}-l_{n+1,n+1}]\nn\\
 &\times&\left(\prod_{k=1}^{n-1} [  l_{k,n-1}-l_{i,n+1} ]
\over \prod_{k\neq i=1}^n [  l_{k,n+1}-l_{i,n+1} ]
\right)^{1/2} |m)_{-in},\label{fn}\\
e_p|m)&=&\sum_{i=1}^n \theta_{ip} (-1)^{\theta_{1p}+ \ldots
+\theta_{i-1,p}+\theta_{i+1,p-1}+ \ldots +\theta_{n,p-1} }
(1-\theta_{i,p-1})\nn\\
&\times&\prod_{k\neq i=1}^n \left(
[l_{i,p+1}-l_{kp}]  [l_{i,p+1}-l_{kp}-1] \over
[l_{i,p+1}-l_{k,p+1}][l_{i,p+1}-l_{k,p-1}-1] \right)^{1/2}|m)_{ip}\nn\\
 &+&\sum_{s=n+1}^p \left( - {\prod_{q=n+1}^{p-1}[l_{q,p-1}-l_{sp}+1]
 \prod_{q=n+1}^{p+1}[l_{q,p+1}-l_{sp}] \over \prod_{q\ne s=n+1}^p
 [l_{qp}-l_{sp}][l_{qp}-l_{sp}+1] } \right)^{1/2} \nn\\
&\times& \prod_{k=1}^n {[l_{kp}-l_{sp}] [l_{kp}-l_{sp}+1] \over
[l_{k,p+1}-l_{sp}][l_{k,p-1}-l_{sp}+1]} |m)_{sp},
\quad (n+1\leq p \leq r-1),\label{ep}\\
f_p|m)&=&\sum_{i=1}^n \theta_{i,p-1} (-1)^{\theta_{1p}+ \ldots
+\theta_{i-1,p}+\theta_{i+1,p-1}+ \ldots +\theta_{n,p-1} }(1-\theta_{ip})\nn\\
&\times&\prod_{k\neq i=1}^n \left(
[l_{i,p+1}-l_{kp}]  [l_{i,p+1}-l_{kp}-1] \over
[l_{i,p+1}-l_{k,p+1}][l_{i,p+1}-l_{k,p-1}-1] \right)^{1/2}
 \nn\\
 &\times& {\prod_{q=n+1}^{p-1}[l_{i,p+1}-l_{q,p-1}-1]
\prod_{q=n+1}^{p+1}[l_{i,p+1}-l_{q,p+1}] \over \prod_{q=n+1}^p
 [l_{i,p+1}-l_{qp}-1][l_{i,p+1}-l_{qp}] }|m)_{-ip} \nn\\
 &+&
 \sum_{s=n+1}^p \left( -
 {\prod_{q=n+1}^{p-1} [l_{q,p-1}-l_{sp}] \prod_{q=n+1}^{p+1}
 [l_{q,p+1}-l_{sp}-1] \over \prod_{q\ne s=n+1}^p
[l_{qp}-l_{sp}-1][l_{qp}-l_{sp}] }\right)^{1/2} |m)_{-sp}, \nn\\
& &\qquad\qquad (n+1\leq p \leq r-1). \label{fp}
\eea
In the above expressions, $\sum_{k\ne i=1}^n$ or
$\prod_{k\ne i=1}^n$ means that
$k$ takes all values from $1$ to $n$ with $k\ne i$. If a vector
from the rhs of (\ref{ki}--\ref{fp}) does not belong to the module
under consideration, then the corresponding term is zero even if the
coefficient in front is undefined; if an equal number of factors in
numerator and denominator are simultaneously equal to zero, they
should be cancelled out. The eqs.~(\ref{ek},\ref{fk}) are the same
as in~\cite{jimbo}; they desribe the transformation of the basis
under the action of the $gl(n)$ generators.

To conclude this section, we shall make a number of comments
on the proof of this Theorem. Provided that all coefficients in
(\ref{ki}--\ref{fp}) are well defined (which is indeed
the case under the conditions required here), it is sufficient
to show that the explicit actions (\ref{ki}--\ref{fp}) satisfy
the relation (\ref{kk}--\ref{eeee}) (plus, of course, also
the $f$-Serre relations). The irreducibility then follows from
the results of Zhang~\cite{zhang} or from the observation that
for generic $q$ a deformed matrix element in the GZ basis is zero
only if the corresponding non-deformed matrix element vanishes.

To show that (\ref{kk}), (\ref{kek}) and (\ref{eifj})
are satisfied is a straighforward matter. The difficult Cartan-Kac
relations to be verified are (\ref{eifi}) and (\ref{enfn}).
We shall consider one case in more detail, namely (\ref{enfn}).
This relation, with the actions (\ref{ki}--\ref{fp}), is valid
if and only if
\bea
&&\sum_{i=1}^n [l_{i,n+1}-l_{n+1,n+1}] {\prod_{k=1}^{n-1}
[l_{k,n-1}-l_{i,n+1}] \over \prod_{k\ne i=1}^n [l_{k,n+1}-l_{i,n+1}]}\nn\\
&&\qquad = \left[ \sum_{k=1}^{n-1}(l_{k,n+1}-l_{k,n-1})+l_{n,n+1}
-l_{n+1,n+1}\right] .
\eea
Putting $a_i=l_{i,n+1}$ for $i=1,2,\ldots,n$, $b_i=l_{i,n-1}$ for
$i=1,2,\ldots,n-1$ and $b_n=l_{n+1,n+1}$,
the identity between $q$-numbers to be proved reduces to
\beq
\sum_{i=1}^n {\prod_{k=1}^n [a_i-b_k] \over \prod_{k\ne i=1}^n
 [a_i-a_k]} = \left[\sum_{k=1}^n (a_k-b_k)\right].
\label{id1}
\eeq
Using the explicit definition of a $q$-number, and relabelling
$q^{2a_i}=A_i$ and $q^{2b_i}=B_i$, this becomes
\beq
\sum_{i=1}^n {\prod_{k=1}^n (A_i-B_k) \over A_i\prod_{k\ne i=1}^n
 (A_i-A_k)} = 1-{B_1B_2\cdots B_n\over A_1A_2\cdots A_n}.
\label{id1q}
\eeq
To prove this last identity, consider the complex function
\beq
f(z)={\prod_{k=1}^n (z-B_k) \over z\prod_{k=1}^n (z-A_k)}. \label{f}
\eeq
This function is holomorphic over $\C$ except in its singular
poles $0$, $A_1,\ldots,A_n$ (under the present conditions, all $A_k$
are indeed distinct). Let $C$ be a closed curve whose interior contains
all these poles. Then the Residue Theorem of complex analysis
implies that
$\oint_C f(z)dz=2\pi i(\Res(0)+\sum_{i=1}^n \Res(A_i))$. It is easy
to see that $\Res(0)=\lim_{z\rightarrow 0} f(z)z = (B_1\cdots B_n)/
(A_1\cdots A_n)$ and that $\Res(A_i)=\lim_{z\rightarrow A_i}
f(z)(z-A_i) = \prod_{k=1}^n (A_i-B_k)/ (A_i\prod_{k\ne i=1}^n
(A_i-A_k))$. On the other hand,
\[
\oint_C f(z)dz = -2\pi i \Res(\infty) = -2\pi i \lim_{z\rightarrow
   \infty} (-z)f(z) = 2\pi i,
\]
and hence the identity~(\ref{id1q}) holds.

For the other cases, the method of proof is similar and we shall
no longer mention the details. For~(\ref{eifi}) with $i=k\leq n-1$,
the identity to be verified is of the following type~:
\beq
\sum_{i=1}^k \left({\prod_{j=1}^k [a_i-b_j][a_i-c_j-1]
\over \prod_{j\ne i=1}^k [a_i-a_j][a_i-a_j-1] } -
{\prod_{j=1}^k [a_i-c_j][a_i-b_j+1] \over \prod_{j\ne i=1}^k [a_i-a_j]
[a_i-a_j+1] }\right) = \left[\sum_{j=1}^k (b_j+c_j-2a_j)\right],
\label{id2}
\eeq
or, using a similar transformation as before,
\bea
&{\ds \sum_{i=1}^k} &\left({\prod_{j=1}^k (A_i-B_j)(A_i-q^2C_j)
\over A_i\prod_{j\ne i=1}^k (A_i-A_j) \prod_{j=1}^k (A_i-q^2A_j) } +
{\prod_{j=1}^k (A_i-C_j)(A_i-q^{-2}B_j) \over A_i\prod_{j\ne i=1}^k (A_i-A_j)
\prod_{j=1}^k (A_i-q^{-2}A_j) }\right) \nn\\
&=& 1-{\prod_{j=1}^k (B_jC_j) \over \left(\prod_{j=1}^k A_j\right)^2}.
\label{id2q}
\eea
This identity is proven by taking the function $f(z)=\prod_j(z-B_j)(z-q^2C_j)/
(z\prod_j(z-A_j)(z-q^2A_j))$ and applying the same Residue theorem.
Finally, the most complicated case is (\ref{eifi}) with $i=p>n$.
The identity to prove is (with $s=p-n$)~:
\bea
&{\ds -\sum_{i=1}^s}& {\prod_{j=1}^s[a_i-c_j][a_i-b_j+1]\over
\prod_{j\ne i=1}^s [a_i-a_j][a_i-a_j+1]} \prod_{k=1}^n{[a_i-d_k-f_k+1]\over
[a_i-d_k+1]} \nn\\
&{\ds +\sum_{i=1}^s}& {\prod_{j=1}^s[a_i-b_j][a_i-c_j-1]\over
\prod_{j\ne i=1}^s [a_i-a_j][a_i-a_j-1]} \prod_{k=1}^n{[a_i-d_k-f_k]\over
[a_i-d_k]} \nn\\
&{\ds +\sum_{k=1}^n}& (f_k){\prod_{j=1}^s[d_k-c_j-1][d_k-b_j]\over
\prod_{j=1}^s[d_k-a_j-1][d_k-a_j]} \prod_{l\ne k=1}^n {[d_k-d_l-f_l]\over
[d_k-d_l]} \nn\\
& =& \left[\sum_{k=1}^n f_k + \sum_{j=1}^s (b_j+c_j-2a_j)\right].
\label{id3}
\eea
Herein, the $f_k$ are equal to $\th_{k,p-1}-\th_{kp}$, and since
the $\th$'s take only the values 0 and 1, the $f_k$'s take only the
values $0,\pm 1$. To prove (\ref{id3}), one again has to use the
same technique on a function
$$ f(z)={\prod_{j=1}^s (z-B_j)(z-q^2C_j) \over z\prod_{j=1}^s
(z-A_j)(z-q^2A_j)}\prod_{k=1}^n{(z-F_kD_k)\over(z-D_k)}.$$
However, it turns out that (\ref{id3}) is true as a general
identity only when in the third summation the factor $(f_k)$ is
replaced by $[f_k]$. In the present case, this can be done without
harm since for the values $x=0,\pm 1$ we have indeed that $[x]=x$.
This completes the verification of the Cartan-Kac relations.

For the $e$-Serre relations, the calculations are extremely lengthy,
but when collecting terms with the same Gel'fand-Zetlin basis
vector and then taking apart the common factors, the remaining
factor always reduces to a simple finite expression which is
easily verified to be zero. These expressions always reduce to one
of the following (trivial) identities~:
\bea
&&[a][b+1]-[a+1][b]=[a-b],\label{38}\\
&&[a+1]+[a-1]=[2][a],\label{39}\\
&&{1\over [a-1][a]}+{1\over[a][a+1]}={[2]\over[a-1][a+1]}.
\eea
In fact, the last of these reduces to the second one, and in some
sense the only identities needed to prove the $e$-Serre relations
are~(\ref{38}) and~(\ref{39}), and combinations of them.
Finally, the calculations for the $f$-Serre relations are of
a similar nature as those for the $e$-Serre relations.

\section{Comments}

We have studied the class of essentially typical representations
of the quantum superalgebra $U_q[gl(n/m)]$ and connected the
relations to be satisfied for these representations with certain
$q$-identities. At present, we do not know how to extend the
present results to other finite-dimensional representations
of $U_q[gl(n/m)]$. In fact, also in the non-deformed case the
problem of how to modify the classical analogs of (\ref{ki}--\ref{fp})
remains an open problem. There is some indication that for
a typical representation the only modification would be to
simply delete those terms for which the coefficient becomes
undefined; however, this is still under investigation and we
hope to report results in the future. For atypical representations,
the GZ basis will presumably be no longer appropriate~: if one
still uses the same GZ-patterns in the case that $[m]$ is atypical,
it turns out that some $|m\rangle$-vectors have a non-trivial
projection both on the maximal submodule and on the quotient
module (of the module spanned by the GZ basis vectors with
a modified action when the corresponding coefficient is
undefined). This property was observed for atypical
representations of $gl(2/2)$, and here again further
investigations are under way.
Note that we also have not examined
the representation theory of $U_q[gl(n/m)]$ in the case of
$q$ being a root of unity.

\vspace{1cm}
{\it Acknowledgements}.
Two of us (T.D.P. and N.I.S.) are grateful to Prof.~Vanden Berghe
for the possibility to work at the Department of Applied Mathematics
and Computer Science, University of Ghent. Prof.~A.\ Salam is acknowledged
for the kind hospitality offered to T.D.P. at the International Center
for Theoretical Physics, Trieste. We would like to thank
Dr.\ Hans De Meyer (Ghent) for stimulating discussions.

This work was supported by the European Community, contract No.\
ERB-CIPA-CT92-2011 (Cooperation in Science and Technology with Central
and Eastern European Countries) and Grant $\Phi$-215 of the
Bulgarian Foundation for Scientific Research.

\addtolength{\baselineskip}{-1mm}

\end{document}